\documentclass[letterpaper,twocolumn,10pt]{article}
\usepackage{usenix,epsfig,endnotes}
\usepackage{multirow}
\usepackage{fixltx2e}
\usepackage{amssymb,amsmath}
\usepackage{hyperref}
\usepackage{url}

\newtheorem{definition}{Definition}

\begin{document}

\date{}

\title{Detecting Spammers via Aggregated Historical Data Set}

\author{
{\rm Eitan Menahem and Rami Puzis}\\
Deutsche-Telekom Laboratories and\\
Information Science Engineering Dept.,\\
Ben-Gurion-University of the Negev\\
Be'er Sheva, 84105, Israel
} 
\maketitle

\thispagestyle{empty}

\begin{abstract}
The battle between email service providers and senders of mass unsolicited emails (Spam) continues to gain traction.
Vast numbers of Spam emails are sent mainly from automatic botnets distributed over the world. 
One method for mitigating Spam in a computationally efficient manner is fast and accurate blacklisting of the senders.
In this work we propose a new sender reputation mechanism that is based on an aggregated historical data-set which encodes the behavior of mail transfer agents over time.
A historical data-set is created from labeled logs of received emails. 
We use machine learning algorithms to build a model that predicts the \emph{spammingness} of mail transfer agents in the near future. 
The proposed mechanism is targeted mainly at large enterprises and email service providers and can be used for updating both the black and the white lists. We evaluate the proposed mechanism using 9.5M anonymized log entries obtained from the biggest Internet service provider in Europe.
Experiments show that proposed method detects more than 94\% of the Spam emails that escaped the blacklist (i.e., TPR), while having less than 0.5\% false-alarms. Therefore, the effectiveness of the proposed method is much higher than of previously reported reputation mechanisms, which rely on emails logs.
In addition, the proposed method, when used for updating both the black and white lists, eliminated the need in automatic content inspection of 4 out of 5 incoming emails, which resulted in dramatic reduction in the filtering computational load.
\end{abstract}




\section{Introduction}
Surveys show that in recent years 76\% to 90\% of all email traffic can be considered abusive \cite{survey3, survey1}. 
A major portion of those billions of Spam emails annually are automatically produced by botnets \cite{behavior1, survey2}. 
Bots create and exploit free web-mail accounts or deliver Spam emails directly to victim mailboxes by exploiting the computational power and network bandwidth of their hosts and sometimes even user credentials. 
Many Spam mitigation methods are used by email service providers and organizations to protect mail boxes of their customers and employees respectively. 
There are three main approaches for Spam mitigation; content-based filtering (CBF), real-time blacklisting or DNS based blacklists, and sender reputation mechanisms (SRM).
All three approaches are briefly described in Section \ref{sec:background}. 

While CBF are considered as the most accurate Spam mitigation methods, they are also the most computationally intensive and sometimes considered as privacy infringing.
In order to speed up the filtering process, organizations maintain blacklists of repeated Spam senders \cite{rbl1,rbl2,rbl3}.
Those blacklists usually complement existing CBF methods by performing the first rough filtering of incoming emails. 
Organizations that choose to maintain their own blacklists gain flexibility in blocking / unblocking certain addresses and the ability to respond to emerging Spam attacks in real-time. 
Flexibility in managing the blacklist is also very important for large email service providers that must react immediately if they receive complaints about emails not reaching their destination.

Sender reputation mechanisms are used to refine blacklisting strategies by learning the liability of mail transfer agents (MTA). 
Most work in this research field focuses on extracting meaningful features from communication patterns of MTAs and from the social network structure \cite{rep1,rep2,rep3,rep4-sn,rep5-sn,rep6,rep7}. 
In Section \ref{sec:related} we elaborate on sender reputation mechanisms.
Beside estimating the liability of MTAs, SRMs also help to respond to emerging Spam attacks in a timely manner.
By analyzing MTA behavior they try to identify patterns which indicate that the particular network address or a subnet is exploited by Spammers. 
SRMs are especially important when one requires a quick detection of sending pattern changes of mail transfer agents. 
For example, a white-listed address belonging to a small bankrupted firm that once maintained legitimate email servers is a lickerish target for exploitation by Spammers. 
A Good SRM should be able to detect a change in the behavior of such address and suggest removing it from a white list and adding it to a black list.

In this paper we investigate methods for updating the reputation of sender MTAs from the perspective of a single email service provider. 
Based on an anonymized log of 9.5M emails, obtained from a large email service provider, we created a historical data set (HDS) that encodes the behavior of sender MTAs over time, as described in Section \ref{sec:hds}. 
Machine learning (ML) algorithms were applied on both the features extracted from the email log and on the HDS in order to create a sender behavior models for deducing the black and white lists. 
We empirically compare email log (EL) based models, HDS based models, and commonly used blacklisting heuristic in a setup where these SRMs are applied on emails that have passed the provider's blacklist.

Based on experiment results, described in Section \ref{sec:results}, we show that an analysis of past behavior of a sender MTAs encoded in HDS enables the filtering out of as much as 94\% Spam emails that have passed the blacklist while having only 0.5\% false positives.
Finally, we also show that by frequent updating of both the black and white-list it is possible to spare content analysis of roughly 82\% of incoming emails.
We discuss the results and limitations of this study and propose directions for future research in Section \ref{sec:conclusions}.

\section{Background}
\label{sec:background}

\begin{table}[t]
\centering
\caption{\label{tab:related} Spam mitigation techniques.}
\begin{tabular}{|l p{2.5cm}|l|}
\hline
CBF & & \cite{cbf1,cbf3,cbf2,cbf4,cbf5} \\
\hline
RBL / DNSBL & & \cite{rbl3,rbl2,rbl1} \\
\hline
\multirow{3}{*}{SRM} 
 & Network level  & \cite{behavior1,rep3,behavior2} \\
 & Transport level & \cite{rep7} \\
 & Application level & \cite{rep4-sn, rep5-sn, rep6, rep-oracle-based} \\
 & Spatio-temporal & \cite{rep2, rep-spatio-temp, rep-predictive-bl} \\
\hline
\end{tabular}
\end{table}

In this section we discuss in further details three prominent approaches for Spam mitigation and some important previous works focusing on machine learning and sender reputation mechanisms. 
Spam-blocking approaches can be roughly divided into three main categories: 
(1) content-based filtering (CBF), (2) real-time blacklisting (RBL), and (3) sender reputation mechanisms (SRM) (see Table \ref{tab:related}). 
Next, we briefly describe the different categories. 

\textbf{Content-based filtering (CBF)} refers to techniques in which emails' body, attached executables, pictures or other files are analyses and processed for producing some features upon which email classification is made \cite{cbf1,cbf2,cbf3,cbf4,cbf5}. 
The email's content is related to the application-level, the highest level of the Open Systems Interconnection (OSI) model. 
Content-based features have a lot of useful information for classification, however, in an Internet Service Provider's (ISP) perspective, there are some disadvantages. First, in order to classify the incoming emails, each email must be put through a relatively heavy-weight content-based filter. 
This leads to a lot of computational resources wasted on filtering, and thus makes it fairly costly when compare to other approaches, such as real-time blacklisting, which will be discussed later.
A second disadvantage of CBF is that Spammers continuously improve their CBF evading techniques. 
For example, instead of sending plain textual Spam emails, they use Spam-images or smarter textual content that obfuscates the unwanted content.

\textbf{Real-time blacklists (RBL)} are IP-based lists that contain IP prefixes of spamming MTAs and are regarded as network-level filters. 
Using the RBL, large firms such as ISPs can filter out emails originating from spamming IPs. 
The filtering is very fast since the decision to accept or reject the email does not require receiving the full email (saving network resources) nor processing its content (saving computational resources). 
In order to avoid misclassification, RBLs must be updated systematically. For example, Spamhaus \cite{rbl1}, Spam-Cop \cite{rbl2}, and SORBS \cite{rbl3} are some initiatives that keep RBL systems updated by tracking and reporting spammer IPs.
RBL methods, however, cannot solve the Spam email problem entirely, as spammers can escape them by, for example, repeatedly changing their IPs by stealing local network IPs \cite{survey2}, or by temporarily stealing IPs using BGP hijacking attacks \cite{behavior1}. 
Another shortcoming of RBL is that whenever an IP prefix is blacklisted, both spammers and benign senders who share the same prefix might be rejected. 
Benign senders can also be blocked because of inaccurate blacklisting heuristics. 
In order to lower the false-positive rates, blacklisting heuristics limit their true positive rates, allowing many Spam-mails to pass the filter and block mainly repeated spammers. 
RBL usually has lower accuracy than CBF, which is an acceptable trade-off given its real-time nature and low utilization of computational resources. 

\textbf{Sender reputation mechanisms (SRM)} for Spam mitigation are methods for computing a liability score for email senders. 
The computation is usually based on information extracted from the network or transport level, social network information, or other useful information sources. 
According to Alperovitch et. al \cite{rep1}, sender reputation systems should react quickly to changes in sender`s behavioral patterns. 
That is, when sender`s sending patterns take a shape of a spammer, his reputation should decrease. 
If the reputation of a sender is below the specified threshold, the system should reject his mails, at least until the sender gains up some reputation by changing his sending properties.
One of the strong advantages of sender reputation mechanisms is that they complement and improve RBLs both in terms of detection accuracy and response to changes. 

Despite the fact that the addresses of MTS which were spotted repeatedly sending Spam will usually not start sending legitimate emails all of a studden, there are several reasons to have accurate SRMs that react quickly to changes in sender behavior. 
First, addresses of once legitimate email servers that are no longer used are exploited by spammers due to their high reputation in databases of large email service providers. 
Spammers can use the window of opportunity created by such addresses for as long as it takes for SRMs to detect the change in behavior of these addresses. 
Quick reaction of SRM is also required when small legitimate email service providers are used to launch massive Spam attacks. 
When such an attack is detected, operations may decide to temporarily blacklist the provider in order to avoid being overwhelmed by Spam emails. 
However, as soon as the attack is over, the provider addresses should be removed from the blacklist.

Finally, the importance of SRMs will further increase with the prevalence of IPv6 in the Internet. 
The impact of botnets on the prevalence of Spam is significantly reduced by blacklisting all known dynamic IP ranges. 
This simple heuristic assumes that legitimate MTAs are not hosted by end users, which are given dynamic IPs by their Internet service providers. 
Currently, in order to simplify the DHCP configuration, dynamic IPv4 addresses are arranged in continuous ranges which are very easy to blacklist.
However, with IPv6, there will not be a necessity for dynamic addresses and the whole address space is expected to become more fragmented, making it difficult to blacklist large IP ranges with simple heuristics.
Furthermore, the extremely large address space and auto-configuration functionality of IPv6 are expected to increase the cost of Spam mitigation \cite{ipv6-spam}. 

Current paper describes SRM that is based on aggregated spatio-temporal and application level features extracted from logs of incoming email.

\section{Related Works}
\label{sec:related}

In this section we discuss some sender reputation mechanisms that share a similar problem domain as the proposed HDS based method.
Several works have used machine learning algorithms to compute sender reputation from data sets which are not based on email content analysis. 

Ramachandran et al. \cite{behavior2}, research a sender reputation and blacklisting approach. 
They present a blacklisting system, SpamTracker, to classify email senders based on their sending behavior rather than the MTA's IP addresses. 
The authors assume that spammers abuse multiple benign MTAs at different times, but their sending patterns tend to remain mostly similar even when shifting from one abused IP to another. 
SpamTracker extracts network-level features, such as received time, remote IP, targeted domain, whether rejected, and uses spectrum clustering algorithms to cluster email servers. 
The clustering is performed on the email destination domains. 
The authors reported a 10.4\% TPR when using SpamTracker on a data set of 620 Spam mails which were missed by the organization filter and eventually were blacklisted in the following weeks.

Tang et al. \cite{rep3} addressed the Spam imbalanced classification task with a new version of SVM, the GSVM-BA. 
The imbalance problem exists in the Spam detection domain due to the fact that there are around 10 Spam mails for each non-Spam mail \cite{survey1}. 
The main two attributes GSVM-BA adds to SVM are a granular computing, which makes it more computationally efficient and a mechanism for under sampling the data set positive instances, so that a good classifier could be learned in a highly unbalanced domain. 
The authors use the proposed GSVN-BA to classify IPs into Spam-IP or non-Spam-IPs, (i.e., learns IPs reputation). 
They used two types of aggregate features, which are both derived from sender's IP, receiver's IPs, and a sending time. 
Their experiments showed a very high precision rate at 99.87\% with a recall of 47\%.

Sender reputation mechanisms are not only limited to network-level features; reputation can also be learned from a much higher communication level, such as the social network level.
The social-network-based filtering approach takes advantage of the natural trust system of social networks.  
For example, if the sender and the receiver belong to the same social group, the communication is probably legitimate. 
On the other hand, if the sender does not belong to any trustworthy social group, he is more likely to be blacklisted.
There are many methods which make a good use of social networks to create both black and white lists. 
For example, J. Balthrop et al. \cite{rep4-sn}, used email address books to compute sender trust degree.

Boykin and Roychowdhury \cite{rep5-sn} extract social network's features from the email header fields such as From, To, and CC, from which they construct a social graph, as can be observed from a single user's perspective. Later, they find cluster of users who can be regarded as trusted. 
Finally, they train a classifier on the email addresses in the white list and black list. 
The authors reported 56\% TPR with the black list and 34\% TPR with the white list. 
Their method, empirically tested on three data sets of several thousands of emails, did not have any false positives.
The downside of the proposed algorithm is that both the black and white lists can be made erroneous. 
For example, the black list can be fooled by attackers which use spyware to learn the user's frequently used address list and have one or more of them added to the Spam mail so that the co-recipients (the Spam victims) will look like they belong to the user's social network. 
The white list can also be fooled by sending Spam mail by using one or more of the user's friends accounts. 
This can be done, for example, if the friend's computer has been compromised by a bot which selectively send Spam mails.

A supervised collaborative approach for learning the reputation of networks is presented by Golbeck and Hendler \cite{rep6}. 
The proposed mechanism is aided by the user's own scores for email senders.  For example, a user can assign a high reputation score to his closest friends, and they in their turn may assign a high reputation rank to their friends. In this way, a reputation network is created. The reputation network may be used as a white-list as a recommendation system with very low false positive rate or allow the user to sort emails by reputation score. 
Similar approach is presented by Xie and Wang \cite{rep-collaborative-2}. They focus on collaboration among several email domains in order to increase the coverage of senders. Each provider compares the email histories obtained from its peers via the proposed Simple Email Reputation Protocol (SERP) with its own records in order to establish trustworthiness of the received data. 


Beverly and Sollins \cite{rep7} investigated the effectiveness of using transport-level features, i.e., round trip time, FIN count, Jitter, and several more.
The best features were selected using the forward fitting method to train a SVM-based classifier. 
They reported 90\% accuracy on 60 training examples. 
However, one of the weaknesses of their method when compared to RBL and other reputation mechanisms is that emails (both Spam and Ham) must be fully received in order to extract their features, thus making the Spam mitigation process less effective.

In addition to the above mentioned SRMs, there is another line of research that focuses on inferring the reputation of senders from spatial and temporal features extracted from emails logs.

SNARE, by Hao et al. \cite{rep2} presented a method based solely on network-level and geodesic features, such as distance in IP space to other email senders or the geographic distance between sender and receiver. 
SNARE classification accuracy has been shown to be comparable with existing static IP blacklists on data extracted from McAfee's TrustedSource system \cite{TrustedSource}.

Liu \cite{rep-oracle-based} proposed a policy that assigned reputation to senders according to the results returned by a contend based filter. 
The author proposes several simple rules that, nevertheless, are able to reduce the number of non caught low volume spammers and accurately selects the set of mixed senders such that the ham/Spam ratio is maximized.  
Finally, West et al. \cite{rep-spatio-temp} introduce a reputation model named PRESTA
(Preventive Spatio-Temporal Aggregation) that is used to predict behavior potential spam senders based on email logs. 
The authors report that PreSTA identifies up to 50\% spam emails that have passed the blacklist having 0.66\% false positives. 

In the current work, we report more than 94\% true positive detection rate and up to 0.6\% false positives in a similar scenario with one week of email logs obtained from a large email service provider.

\section{Learning From the Email Log}	
\label{sec:el}

The data set used in this research contained 168 hours (7 days) of anonymized email log obtained from a large email service provider. 
The provider maintains its own Spam mitigation solution which includes black-and white lists and a CBF.
The email log includes only emails that have passed the black-list and are labeled by a CBF named eXpurgate \cite{expurgate}. 
The developers of eXpurgate claim \emph{``A Spam recognition rate of over 99\%''} and \emph{``zero false positives''} with unpublished false negative rate.

The email log was parsed to create a relational data set in the following way.
Let \(IP\) denote the Internet address of the Sender MTA. 
In the following discussions we will assume an implicit partition of the IP address into four fields: two fields (MSB and LSB) of in CIDR (Classless Inter-Domain Routing) notation, the subnet identifiers \(IP/8\), \(IP/16\), \(IP/24\), and the host identifier \(IP/32\).
Let \(EL=\{IP,T,NR,AE,PT,SpamClass\}^m\) be the relational data set representing the email log.
It contains one line for each received mail where \(T\) is the receiving time, \(NR\) is the number of recipients, \(AE\) is the number of addressing errors, \(PT\) is the time spent by eXpurgate for processing the email,  \(SpamClass\) is the binary mail classification (\(Spam=1\) or \(Ham=0\)) obtained from eXpurgate, and \(m\) is the number of emails. 
Table \ref{tab:el-ds} presents the properties of the \(EL\) data set, used in the experiment. 
Table \ref{tab:el-example} is an example of \(EL\).
This example will be used to demonstrate the construction of HDS in Section \ref{sec:hds}.

\begin{table}[t]
\centering
\caption{\label{tab:el-example} Example of an email log}
\begin{tabular}{|c|p{0.5cm} c p{0.5cm} |c|}
\hline
\# & IP & T & AE & Spam Class \\
\hline
1 & IP1 & 1   & 6 & 0 \\
2 & IP1 & 1.5 & 2 & 0 \\
3 & IP1 & 2.8 & 3 & 1 \\
4 & IP1 & 4.1 & 0 & 0 \\
5 & IP1 & 5.5 & 2 & 0 \\
6 & IP1 & 6.3 & 2 & 0 \\
7 & IP1 & 7.1 & 57& 1 \\
8 & IP1 & 7.9 & 48& 1 \\
9 & IP3 & 9   & 53& 1 \\
10& IP2 & 11  & 2 & 0 \\
\hline
\end{tabular}
\end{table}

We used machine learning (ML) algorithms to create sender reputation models based on \(EL\).
As can be seen from Table \ref{tab:el-attr-rank}, the subnet identifiers play significant role in ML models based on \(EL\).
This fact suggests that ML algorithms are able to identify spamming addresses even based on emails that have passed the blacklist of the email service provider.
We will denote the model built by ML from the \(EL\) data set as EL-Based SRM.

\begin{table}[t]
\centering
\footnotesize
\caption{\label{tab:el-attr-rank} Infogain ranking of \(EL\) features.}
\begin{tabular}{|l|@{\hspace{1.1mm}}c@{\hspace{1.1mm}}|@{\hspace{1.1mm}}c@{\hspace{1.1mm}}|@{\hspace{1.1mm}}c@{\hspace{1.1mm}}|@{\hspace{1.1mm}}c@{\hspace{1.1mm}}|@{\hspace{1.1mm}}c@{\hspace{1.1mm}}|@{\hspace{1.1mm}}c@{\hspace{1.1mm}}|@{\hspace{1.1mm}}c@{\hspace{1.1mm}}|@{\hspace{1.1mm}}c@{\hspace{1.1mm}}|}
\hline
\multirow{2}{*}{} & \multicolumn{4}{c|@{\hspace{1.1mm}}}{IP} & \multirow{2}{*}{NR} & \multirow{2}{*}{AE} & \multirow{2}{*}{PT} & \multirow{2}{*}{T} \\
&/8& /16 & /24 & /32 & & & & \\
\hline
Rank & 0.262 & 0.087 &  0.056 & 0.015 & 0.344 & 0.0002 & 0.114 & 0.041 \\
\hline
\end{tabular}
\end{table}

Let \(BLT\) and \(WLT\) be the blacklisting and the whitelisting thresholds, respectively.
If the EL-Based SRM classifies an \(EL\) instance as Spam with confidence above \(BLT\) the respective IP address is added to the black-list.
The white-list is updated symmetrically. 
A similar approach using a different set of features, was taken in \cite{rep2}.
In Section \ref{sec:results} we will evaluate the effectiveness of EL-Based SRM for updating the black and the white-lists and use it as a baseline for comparison to the proposed SRM.
It should be noted that the performance of the EL-Based SRM roughly matches the performance of state-of-the-art methods.

\begin{table}[t]
\centering
\caption{\label{tab:el-ds} The email log data set}
\begin{tabular}{|l|c|}
\hline 
Alias & \(EL\) \\
\hline
Source & email logs \\
\hline
Attributes & 9 \\
\hline
Target attribute & Spam / Ham \\
\hline
Spam instances & 12.25\%\\
\hline
Time frame & 168 hours \\
\hline
Instances & 9,507,154 \\
\hline
Distinct IPs & 678,509 \\
\hline
\end{tabular}
\end{table}

Another simple SRM which can be applied directly on the EL data set is a heuristic that detects repeated spammers and trusted legitimate email senders.
This Heuristic-SRM is based solely on the fraction of Spam mails sent by the \(IP\) in the past.
Let \([T_{start},T_{end})\) denote a continuous time range starting at \(T_{start}\) (inclusive) and ending at \(T_{end}\) (exclusive).
\begin{definition}
\label{def:spammingness}
Spammingness of a sender IP (denoted by \(Y_{IP,[T_{start},T_{end})}\)) is the  fraction of Spam emails sent by the \(IP\) during the time window \([T_{start},T_{end})\).
\end{definition}
If the spammingness of an IP in the past is above \(BLT\), we blacklist it. 
Symmetrically, if the spammingness of an IP is below \(WLT\), we put it in the white-list.
Since most of the IP addresses are either legitimate email senders or potential spammers this heuristic performs quite well in practice for large time windows.
However, it can hardly detect changes in the sender behavior and therefore cannot react to it in a timely manner.

Making aggregations which capture the entire history of an \(IP\) may not be optimal as it releases the focus from the most recent statistics that may indicate behavior changes. 
A better approach is to split the history on an \(IP\) into several non-congruent time windows, as described in the next section. 

\section{Historical Data Set}
\label{sec:hds}

The primary objective of the presented work is to automatically learn a classification model for updating the black and the white lists in a timely manner.
We assume that there are differences in the behavioral patterns of spammers and legitimate IPs and try to capture these differences. 
We also assume that the behavioral patterns of \(IP\)s may change over time due to adversary actions, such as stealing local network IPs of benign MTAs, by temporarily stealing IPs using BGP hijacking attacks, or by installing a bot agent at a end-user's device. For example, a small business that runs its own legitimate email servers may be subverted by some malware and become a Spam sender. 
In this case the IP may temporary enter the blacklists of large email service providers until the problem is mitigated. 
We further assume that an analysis of a long period of time is important for blacklisting repeated spammers. 
However the recent behavior is important for rapid reaction to changes in the sender behavior. 

\begin{figure}[ht]
\centering
\includegraphics[width=3in]{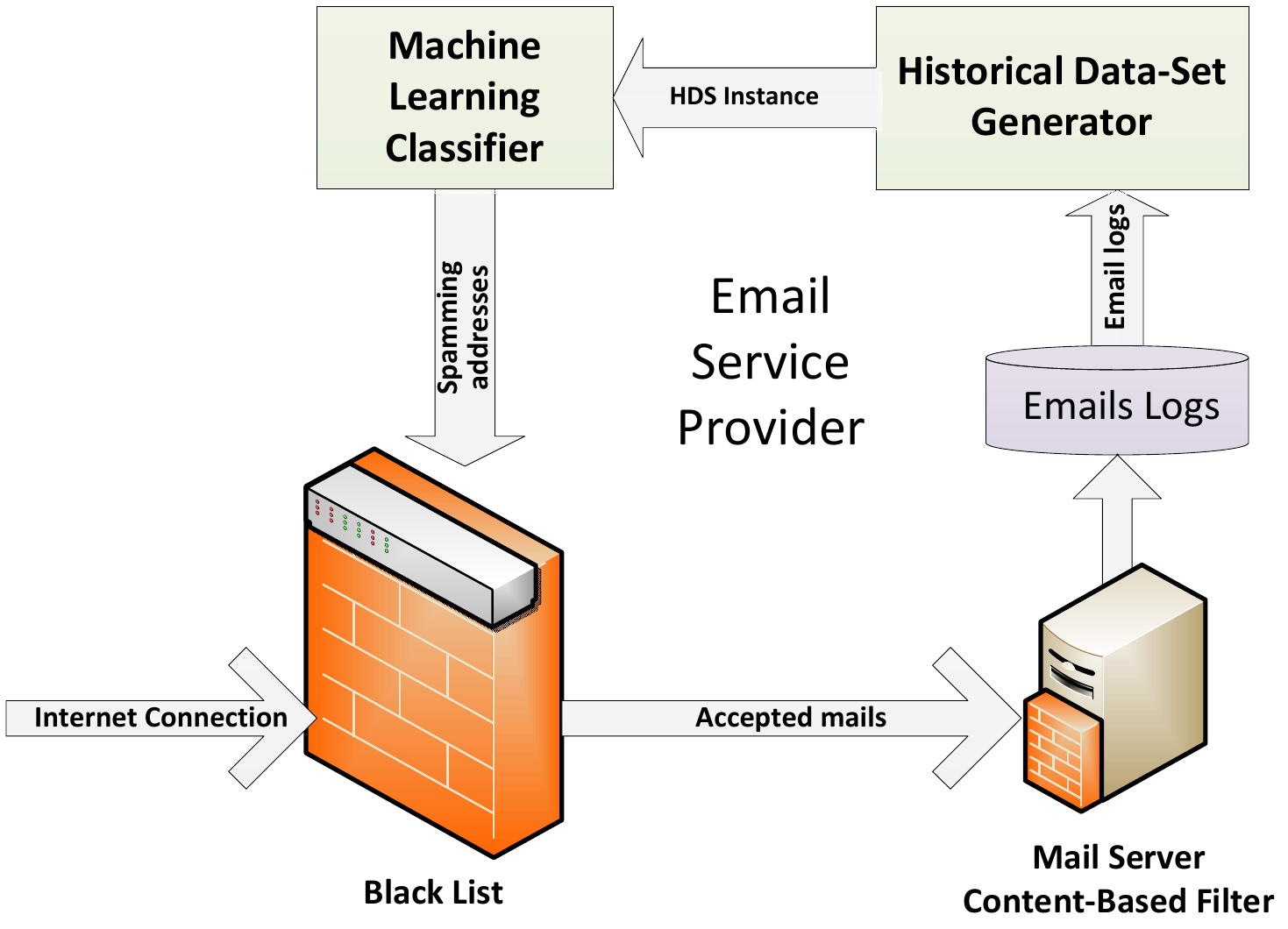}
\caption{The HDS work flow.}
\label{fig:architecture}
\end{figure}

The idea of HDS in a nutshell is to aggregate email log records across multiple variable length historical time windows in order to create informative statistical records. 

For each historical time window, HDS records contain multiple aggregations of the attributes in the \(EL\) data set. 
The statistical learning algorithm is applied on the aggregated features to predict the behavior of an IP in the near future.  
Based on this prediction, HDS-based SRM will temporarily update the black and white lists.

Intuitively, one special property of the proposed method is its ability to model the subject behavior over time (which is more informative than only its current state). In particular, it extracts features for a sender MTA by aggregating past transactions, which are then labeled using its behavior on future transactions. An HDS-based classification model, therefore, can estimate MTA behavior in the near future based on its past transactions. This information allows detection of 'Spammer' behavior patterns even with only very few Spam emails sent. This unique property can speed-up the blacklisting process, and hence improve the TPR.

Next, we define the HDS building blocks. 
Figure \ref{fig:architecture} depicts the general work flow of HDS based SRM.
HDS records are uniquely identified by a reference time \(T_0\) and \(IP\). 
\(EL\) records preceding \(T_0\) are denoted with negative time indexes e.g. \(T_{-1}\). 
An Historical time window is a continuous range \([T_{-w_0\cdot 2^i},T_0)\) where \(w_0\) is the length of the smallest historical time window and \(i\) is a positive integer.
Using exponentially growing historical time windows gives us two benefits. First, we are able to capture a long history of an IP without exploding the size of the HDS records. 
Second, the size of the most recent time windows is small enough to detect the slightest change in the behavior of an IP.
In Section \ref{sec:results} we will show that the number of historical time windows should be carefully chosen in order to obtain the best performance. 
Choosing the best length of the smallest time window (\(w_0\)) is not intuitive either, however, will not be covered in this paper. 

Let \(FS_{IP,T_0,i}\) be a set of aggregated features of a particular \(IP\), computed using \(EL\) records in a historical time window \([T_{w_0\cdot 2^i},T_0)\).
Each HDS record contains \(n\) feature sets (\(FS_{IP,T_0,0}, FS_{IP,T_0,1},\ldots,FS_{IP,T_0,n-1}\)). 
Every feature set \(FS_{IP,T_0,i}\) includes aggregates of all features extracted from the email logs.
The actual set of features depends on the email service provider and the nature of the email logs. 
In this paper we have constructed feature sets for the HDS records by taking the sum, mean, and variance of the number of recipients (\(NR\)), the number of addressing errors (\(AE\)), the CBF processing time (\(PT\)), and the \(SpamClass\).
Note that the mean of \(SpamClass\) is in fact the spammingness of the IP in the respective time window. 
In addition, for each time window we have also included the total number of emails sent and the number of times the sender changed their behavior from sending Spam to sending legitimate emails, and vice versa.
The last feature plays a significant role in the classification of senders as can be seen from Table \ref{tab:hds-feature-rank}.
\begin{definition}
\label{def:erraticness}
Erraticness of a sender \(IP\) (denoted by \(Z_{IP,[T_{start},T_{end})}\)) is the  number times the sender changed his behavior from sending Spam to sending legitimate emails and vice versa during the time window \([T_{start},T_{end})\). 
\end{definition}
\vspace{-0.1cm}
The goal of the \emph{Erraticness} method is to detect MTAs that are about to change behavior to `Spamming' and will stay that way for a long period of time, i.e., have a small Erraticness value. Note that the underlying notion of \emph{Erraticness}, that MTAs sending behavior can alternate, i.e., MTA can send ham emails for a while, then send some Spam emails for a while and later go back to send ham emails once again, is indeed realistic. In fact, our EL-dataset (discussed in section \ref{sec:datasets}) shows that 25.6\%  of the MTAs changed their sending behavior during a single day (8,804 IPs change their behavior more than 4 times). In section \ref{sec:results} we show that the HDS (\emph{Erraticness}) can learn to avoid blacklisting such MTAs to avoid false-positives. ý

\begin{table}[t]
\centering
\caption{\label{tab:hds-feature-rank} Infogain ranking of HDS features (\(n=5\), \(w_0=60min\)).}
\includegraphics [width=3in] {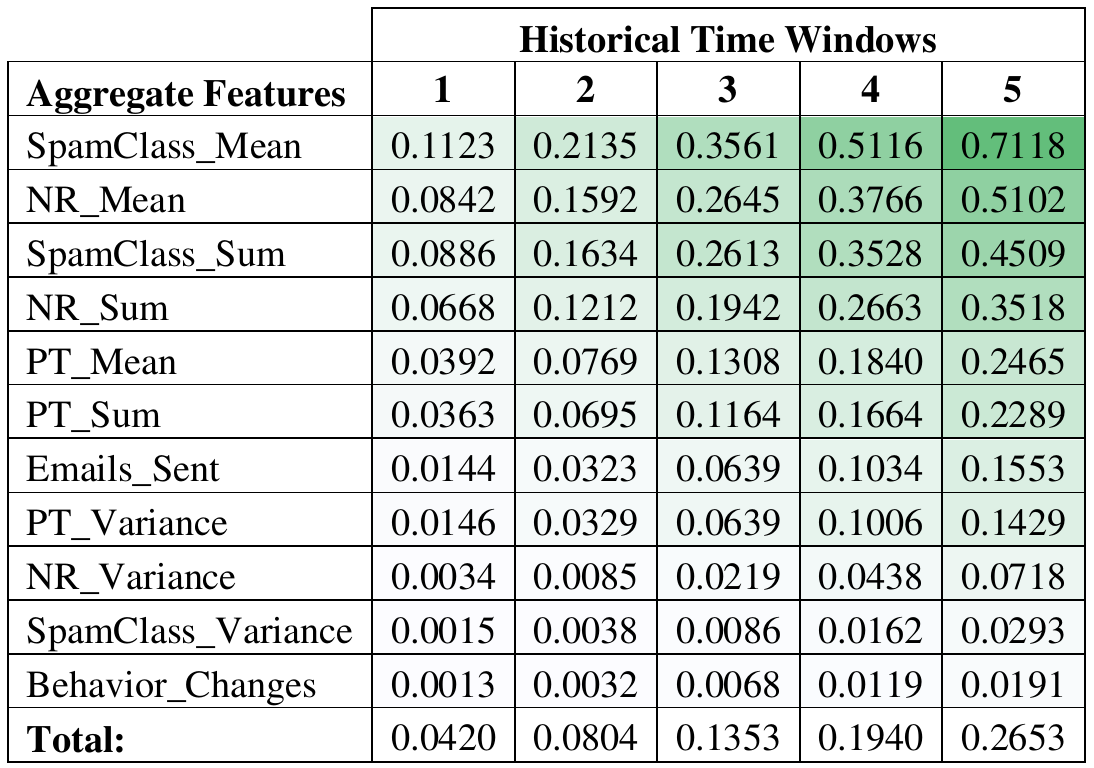}
\end{table}

Let \([T_0,T_{Pred})\) be the prediction time window where \(Pred\) is its length. 
Let \(Class_{IP,[T_{0},T_{Pred})}\) be the target attribute of the HDS records used to train machine learning classifiers.
\(HDS\) records are identified by an \(IP\) and a reference time \(T_0\). 
They contain \(n\) feature sets that correspond to \(n\) historical time windows and a target attribute. 
Let \(HDS\) be a relational data set derived from \(EL\):
\[HDS=(IP,T_0,FS_{IP,T_0,1},\ldots,FS_{IP,T_0,n},Class_{IP,[T_{0},T_{Pred})})^l\]
where \(l\) is the number of records.

Table \ref{tab:hds-struct} depicts the historical data set structure. 
Each HDS record is identified by \(IP\) and a reference time \(T_0\) and contains \(n\) feature sets. 
\begin{table}[t]
\centering
\caption{\label{tab:hds-struct} The HDS structure}
\begin{tabular}{|c|c|c|c|c|p{0.6in}|}
\hline
IP & \(T_0\) & \multicolumn{3}{c|}{Feature Sets} & Emerging Behavior \\
\hline
\(IP_1\)&\(1\)&\(FS_{IP_1,1,1}\)&\ldots&\(FS_{IP_1,1,n}\)&\(Class_{IP_1,1}\)\\
\(IP_2\)&\(1\)&\(FS_{IP_2,1,1}\)&\ldots&\(FS_{IP_2,1,n}\)&\(Class_{IP_2,1}\)\\
\(IP_1\)&\(2\)&\(FS_{IP_1,2,1}\)&\ldots&\(FS_{IP_1,2,n}\)&\(Class_{IP_1,2}\)\\
\(IP_2\)&\(2\)&\(FS_{IP_2,2,1}\)&\ldots&\(FS_{IP_2,2,n}\)&\(Class_{IP_2,2}\)\\
\multicolumn{6}{|l|}{\vdots} \\
\hline
\end{tabular}
\end{table}

Next, we describe two variants of an HDS-Based SRM.
The target attribute of the first variant is the future Spammingness of an IP and the target attribute of the second variant is its future Erraticness.
We will denote these two variants as HDS-Based (Spammingness) SRM and HDS-Based (Erraticness) SRM, respectively.

\begin{table*}[ht]
\centering
\caption{\label{tab:hds-example} Example of an HDS }
\begin{tabular}{|c|c|c|c|c|c|c|c|c|c|c|}
\hline
 & & \multicolumn{2}{c|}{\([T_{-8},T_0)\)} & \multicolumn{2}{c|}{\([T_{-4},T_0)\)} & \multicolumn{2}{c|}{\([T_{-2},T_0)\)} & \multicolumn{2}{c|}{\([T_{-1},T_0)\)} & \([T_0,T_{+4})\) \\
IP & \(T_0\) & EC & AE & EC & AE & EC & AE & EC & AE & Spammingness \\
\hline
\(IP_1\)&2&-&-&-&-&2&8&1&2&0.333\\
\(IP_1\)&4&-&-&3&11&1&3&0&0&0.200\\
\(IP_1\)&6&-&-&3&5&2&2&1&2&0.400\\
\(IP_1\)&8&8&120&5&109&3&107&2&105&0.500\\
\hline
\end{tabular}
\end{table*}

\subsection{HDS-Based (Spammingness) SRM}

In order to train the HDS-Based (Spammingness) classifier, we set the target attribute of every HDS record to be the Spammingness of the \(IP\) in a time period following \(T_0\) (\(Y_{IP,[T_{0},T_{Pred})}\)).
Table \ref{tab:hds-example} shows an example of an historical data set derived from the email log in Table \ref{tab:el-example}. 
HDS contains one instance per IP per time unit, where time unit was defined as the size of the smallest time window (\(w_0\)).
In this example, \(w_0=1\), \(n=4\), and \(T_{Pred}=4\).
\emph{EC} stands for email count and \emph{AE} stands for the sum of addressing errors.
Both are examples of aggregated features that are calculated for each one the four feature sets, while \emph{Spammingness} is calculated according to Definition \ref{def:spammingness}.
In our experiments, each feature set contained 13 different features (see Table \ref{tab:hds-feature-rank}).

Given the HDS with a target attribute set to Spammingness, we can train a machine learning based classifier to predict the future Spammingness of IP addresses.
These predictions are used by the HDS-Based (Spammingness) SRM to roughly distinguish between spammers and non spammers.
If the predicted Spammingness is higher than a given threshold (e.g., 0.5), we then apply the \(BLT\) threshold on Spammingness (i.e \(SpamClass_mean\)) in the largest historical time window.
Symmetrically, if the predicted Spammingness is lower than the given threshold the \(WLT\), is applied to determine whether or not the IP is added to the white-list.

The operation of HDS-Based (Spammingness) SRM resembles the Heuristic SRM described in Section \ref{sec:el}.
Here, however, the heuristic rule is augmented by the prediction of a machine learning classifier. 
In Section \ref{sec:results} we will show that this combination produces a high quality SRM.
Preliminary experiments (not presented in this paper) show that using the predicted Spammingness only, without applying the \(BLT\) and \(WLT\) thresholds on historical Spammingness, results in poor classification.

\subsection{HDS-Based (Erraticness) SRM}

Another variant of the HDS-Based SRM uses a machine learning classifier to predict the stability or Erraticness of the IP behavior in the prediction time window. In order to train the HDS-Based (Erraticness) classifier we set the target attribute of every HDS record to be the Erraticness of the \(IP\) in a time period following \(T_0\) (\(Z_{IP,[T_{0},T_{Pred})}\)).
Classifiers trained on this data are used in a slightly different way than classifiers trained to predict the Spammingness of an IP.
The difference is mainly in the rule we use to arrive at the second part of the black-listing / white-listing decision process.
If the ML model predicts an unvarying behavior (\(Erraticness<<\epsilon\)), after T0, meaning that the IP is not expected to change its behavior in the nearest future, we apply the same rule that guided the heuristic-SRM.

First, we check whether the predicted Erraticness is very close to zero, meaning that \(IP\) is not expected to change its behavior in the nearest future.
If it is, then the same rule that guides the Heuristic-SRM is applied.

That is, if the Spammingness of the IP in the longest historical time window is above the blacklisting threshold \(BLT\), then the IP is blacklisted.
Symmetrically, if changes in the IP behavior are not predicted and the IP Spammingness is below the whitelisting threshold, then the IP is added to the white-list. 


Experiment results presented in Section \ref{sec:results} show that, in terms of AUC score, the HDS-Based (Spammingness) SRM consistently performs better than all the other examined SRMs. In fact, the former SRM receives the highest performance metrics and is the most tolerant to configuration changes.

\section{Evaluation Methodology}
\label{sec:experiments}

\subsection{Evaluation Environment}
\label{sec:eval-env}
It is not possible to directly compare ML models trained on the \(EL\) and the \(HDS\) data sets. 
The main difficulty is that the number of instances and the target-variable in both data sets is different. 
Each \(EL\) instance corresponds to a single email, while each \(HDS\) instance represents email aggregates for a time period. 
Therefore, we implemented a unique evaluation environment around WEKA machine-learning tool \cite{DBLP:journals/sigkdd/HallFHPRW09} that allows for an evaluation of the aforementioned SRMs on a common ground. 
In addition to the HDS-based and the EL-based SRMs, we implemented an Huristic-SRM. The Huristic-SRM blacklists \(IP\)s that, during a past time window, had a spam fraction value greater than the \emph{BLT} value. As opposed to HDS and EL SRMs, the Huristic SRM does not use machine-learning, and therefore, does not need training. The Hueristic-SRM is currently deployed at the ISP, from which our data is originated, and is used for updating their black-list.
The evaluation environment was designed to simulate the general filtering process of incoming emails.

\begin{figure}[ht]
\vspace{-0.5cm} 
\centering
\caption{\label{fig:methodology} The HDS evaluation space.}
\includegraphics[width=2.7in]{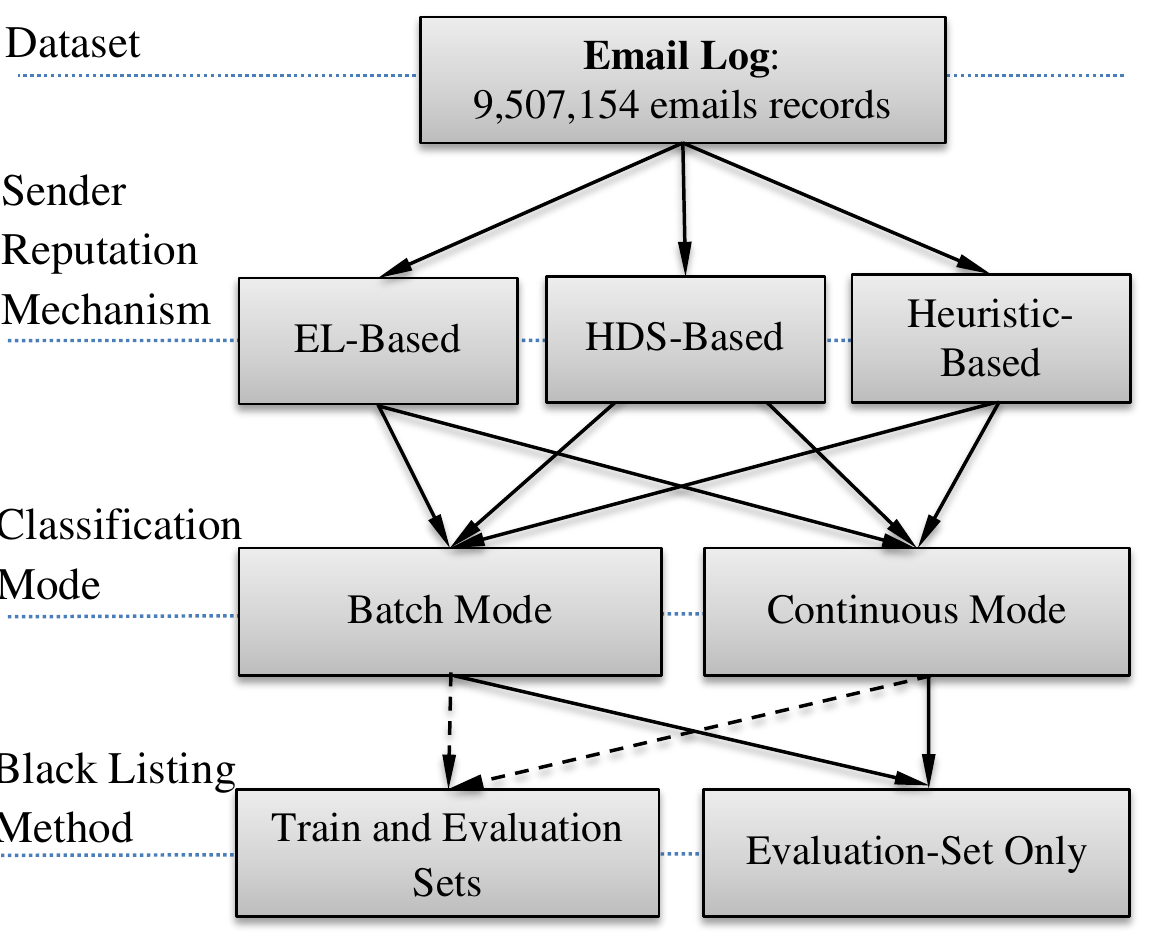}
\vspace{-0.1cm} 
\end{figure}

\begin{figure}[t]
\centering
\includegraphics[width=2.8in]{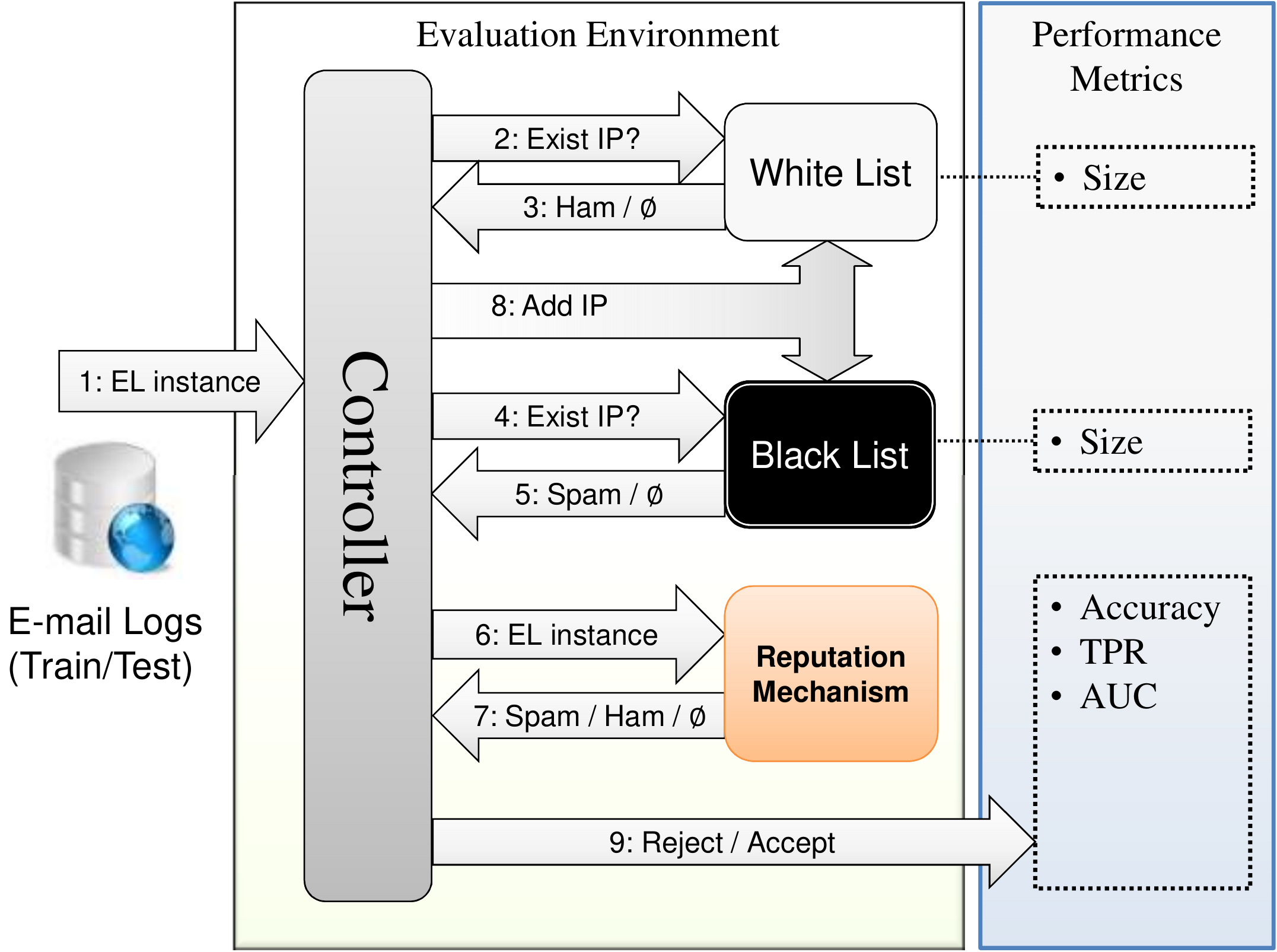}
\caption{\label{fig:exp-architecture} HDS evaluation environment}
\vspace{-0.5cm}
\end{figure}

The HDS evaluation environment, (see Figure \ref{fig:exp-architecture}) contains four modules: controller, while-list, black-list, and a reputation mechanism. 
The sender \(IP\) of every incoming email (i.e., an \(EL\) instance) is first looked up in the white-list (steps 1: and 2: in Figure \ref{fig:exp-architecture}).
A positive result cause the email to be accepted (steps 3: and 9:).
\(EL\) instances that have been accepted are passed to the SRM (step 6:) in order to update the reputation model.
Emails arriving from blacklisted MTAs are rejected without further processing (steps 4:, 5:, and 9:).
If neither the white-list nor the black-list contain the \(IP\), the classification result of the SRM is used to make the final decision and to update the lists, if necessary (steps 6:, 7:, 8:, and 9:).
Note that in the case of the HDS-Based and the Heuristic SRMs, emails received from blacklisted IPs are ignored since they never reach the reputation mechanism. 
The general experimental settings are depicted in Figure \ref{fig:methodology}.

In contrast to continuous classification where each email received from non blacklisted \(IP\) is passed to the SRM, the evaluation environment can operate in a batch mode. In this mode the incoming emails are logged, but the reputation mechanism is activated once in a predefined time period.
After processing the logged emails, the SRM returns the controller two sets of addresses.
One set contains the addresses that should be blacklisted and the other contains the addresses that should be whitelisted. 
Either set, of course, can be empty. 

\begin{table*}[ht]
\centering
\small
\caption{\label{tab:exp-setup} Summary of the experiment setup.}
\begin{tabular}{|l|p{0.9in}|p{0.7in}|p{0.6in}|l@{\hspace{0.9mm}}|@{\hspace{0.9mm}}l@{\hspace{0.9mm}}|@{\hspace{0.9mm}}l@{\hspace{0.9mm}}|p{0.4in}|p{0.2in}|@{\hspace{0.9mm}}l@{\hspace{0.9mm}}|p{0.7in}|}
\hline
Section  & SRM & Machine Learning Algorithm & Batch \newline Frequency & \(w_0\) & \(n\) & \(T_{Pred}\) & History Size & BLT & WLT & Performance \newline Metrics\\
\hline
{\multirow{6}{*}{\ref{sec:results-el-hds}}}
 &  Heuristic & n/a & \multirow{4}{2.8in} {Continuous} & n/a & n/a & n/a & \multirow{3}{0.9mm}{960m} & \multirow{14}{*}{0.5} & \multirow{14}{*}{0.05} & \multirow{9}{0.8in}{\% Error,\newline TPR,\newline FPR,\newline AUC,\newline BL size,\newline WL hits} \\ \cline{2-3} \cline{5-7}
 & HDS-Based (Spammingness) & \multirow{4}{0.8in}{Na\"{i}ve Bayes, \newline C4.5, \newline Logistic Regression, \newline BayesNet}& & \multirow{2}{*}{60m} & \multirow{2}{*}{5} & \multirow{2}{*}{60m} & & & & \\ \cline{2-2}
 & HDS-Based \hspace{10pt}(Erraticness) & & & & & & & & & \\ \cline{2-2} \cline{5-8}
 & EL-Based & & & n/a & n/a & n/a & n/a & & & \\ 
\cline{1-8}
{\multirow{6}{*}{\ref{sec:results-batch}}}
 &  Heuristic & \multirow{4}{0.8in}{BayesNet} & \multirow{4}{0.8in}{1/2 min,\newline 1 min,\newline 2 min,\newline 5 min,\newline 20 min, \newline 60 min} & n/a & n/a & n/a & \multirow{3}{*}{960m} & & & \\ \cline{2-2} \cline{5-7}
 & HDS-Based (Spammingness) & & & \multirow{2}{*}{60m} & \multirow{2}{*}{5} & \multirow{2}{*}{60m} & & & & \\ \cline{2-2}
 & HDS-Based \hspace{10pt}(Erraticness) & & & & & & & & & \\ \cline{2-2} \cline{5-8}
 & EL-Based & & & n/a & n/a & n/a & n/a & & & \\ 
\cline{1-8}
{\multirow{2}{*}{\ref{sec:results-n}}}
 & HDS-Based \hspace{10pt}(Erraticness) & \multirow{1}{0.8in}{BayesNet} & Continuous & 15s & \(1..14\) & 60m & 15s ...\newline 2,048m & & & \\
\hline
\end{tabular}
\end{table*}

\subsection{Dataset}
\label{sec:datasets}
To evaluate the proposed SRMs, we made use of a single email log datasets which contained 9.507 million anonymized log entries (emails headers) of 678,509 distincs IPs, which were received during a 168 hours (7 days) period at T-Online ISP. The dataset is comprised of 9 attribues and 12.25\% `Spam' labeled instances. The un-received emails, which were blocked by the T-Online black-list, were not logged and therefore their headers are not included in the dataset. The dataset was fully labeled by an automatic content-based filtering device, eXpurgate \footnote{Eleven, eXpurgate Anti Spam,  \url{http://www.eleven.de/overview-antispam.html}}. The eXpurgate claims "A Spam recognition rate of over 99\%" and "zero false positives" with unpublished false negative rate. The dataset was partitioned into training and validation sets, each containing the instances of all the emails sent by 200k randomly selected sender IPs. The training set (2,835,214 instances) and validation sets (2,864,208 instances) are mutually exclusive, meaning that $IP$s that exist in the training set do not appear in the validation set, and vice versa.

\subsection{Performance Metrics}

In order to evaluate the sender reputation mechanisms discussed in this paper, we use the following performance metrics: classification error rate, true positive rate, false positive rate, area under the ROC (Receiver Operating Characteristic) curve, black-list size, and number of white-list hits. 
These performance metrics provide enough information to assess how well the HDS-Based SRM could be used to both reduce the load from mail servers, and reduce the number of potential customer complaints.

\emph{Classification error rate} is the rate of incorrect predictions made by a classifier and is computed by the following equation:
\[Error=\frac{FP+FN}{TP+TN+FP+FN}\] 
where TP, TN, FP and FN stand for the number of true positive, true negative, false positive, and false negative rejections of emails respectively.

We will use the \emph{Area Under the ROC Curve (AUC)} measure in the evaluation process. 
The ROC curve is a graph produced by plotting the true positive rate (\(TPR=TP/(TP+FN)\)) versus the false positive rate (\(FPR=FP/(FP+TN)\)).
The AUC value of the best possible classifier will be equal to unity. 
This would imply that it is possible to configure the classifier so that it will have 0\% false positive and 100\% true positive classifications. 
The worst possible binary classifier (obtained by flipping a coin for example) has an AUC of 0.5.
The AUC is considered as an objective performance metric as it does not depend on the specific discrimination threshold used by a classifier. 

\emph{Black list size} is the number of IPs added to the black list during each experiment execution. 
The blacklist size mainly affects the \(IP\) lookup time and the amount of computational resources spent on its maintenance \cite{iplookup}.
Faster lookup times mean less delay in email delivery, while the computational resources required to maintain the black-list directly translate into cost.

The number of \emph{white-list hits} is an indication of the number of emails that were delivered without content inspection.
A Higher number of white-list hits means less computational resources spent on Spam filtering. 

\section{Experimental Results}
\label{sec:results}

In order to assess the effectiveness of HDS-Based SRMs, we implemented HDS-Based (Erraticness), HDS-Based (Spammingness), EL-Based, and Heuristic SRMs within WEKA machine-learning framework \footnote{The code, dataset, and operation instructions can be downloaded from \emph{\#Anonymized\#}} \cite{weka}.
The evaluation environment presented in Figure \ref{fig:exp-architecture} was also implemented within WEKA as a special classifier which uses the SRMs to update the black and white lists. 
The training of the EL-Based and HDS-Based sender reputation algorithms was made using a cross validation process.
The data sets were ordered chronologically to preserve the order in which the emails were received. 
The same cross-validation procedure was applied to all tested settings, shown in Table \ref{tab:exp-setup}.

The blacklisting threshold \(BLT\) and the whitelisting threshold \(WLT\), as described in Sections \ref{sec:el} and \ref{sec:hds}, are parameters of the evaluation environment.
In the following experiments these two threshold were fixed and equal for all SRMs.
The value of \(BLT\) and \(WLT\) were empirically chosen to be \(0.5\) and \(0.05\) respectively.
These values assure low false positive rates while resulting in relatively high true positive rates.

Both the black and the white lists were empty in the beginning of the experiments.
It is also possible to initialize the black and white lists by executing the evaluated SRM on the train data. 
In this scenario the address lists already contain valuable information in the beginning of the testing phase.
In this paper, however, we focus on the more challenging task that involves the construction of the address lists from scratch.

\subsection{The Value of HDS Aggregations} 
\label{sec:results-el-hds}
The first and most important question in this study is whether the HDS-based SRM outperforms other related SRMs. To answer this we compare HDS-based SRM to EL-based SRM on an identical test bed. We argue that in general HDS-based features are at least as informative as EL-based features because they are derived directly from EL-based features and thus HDS features should produce a superior classification model, with respect to that produced using the corresponding EL dataset. In this experiment we simulated a condition in which every email, whose sender's \(IP\) is not included in either the black or white lists, is classified by the tested SRMs on arrival. We denote this classification mode as 'continuous mode'.

In order to increase the reliability of the experimental results, we evaluated the HDS-Based and EL-Based sender reputation mechanisms on four different machine-learning algorithms from separated machine-learning families.
The algorithms used were: Na\"{i}ve Bayes (Bayes) \cite{ml-naiive-bayes}, C4.5 (Decision Trees) \cite{ml-c45}, Logistic Regression (Function) \cite{ml-Logistic-Regression}, and BayesNet (Bayes). Due to memory constraints, in this experiment the train and validation set contained the email sending of 50,000 randomly selected IPs.

The HDS instances were generated using the following parameters: \(w_0=60\) minutes, \(n=5\), and \(T_{Pred}=60\) minutes. 
The total history length is: 
\[\Delta T=w_0\cdot 2^{(n-1)}=960 \hspace{4pt} minutes \hspace{4pt}  (16 \hspace{4pt} hours)\]

Table \ref{tab:res-srm} presents the results of this experiment.
We also applied the Heuristic SRM as a baseline to compare with other techniques.
The time window used by the Heuristic SRM for computing the spammingness of the \(IP\) addresses was set to 960 minutes.
The blacklisting and the whitelisting thresholds were set to \(BLT=0.5\) and \(WLT=0.05\), respectively for all SRMs.

\begin{table}[t]
\centering
\caption{\label{tab:res-srm} Dependability on ML classifiers.}
\includegraphics[width=3.15in]{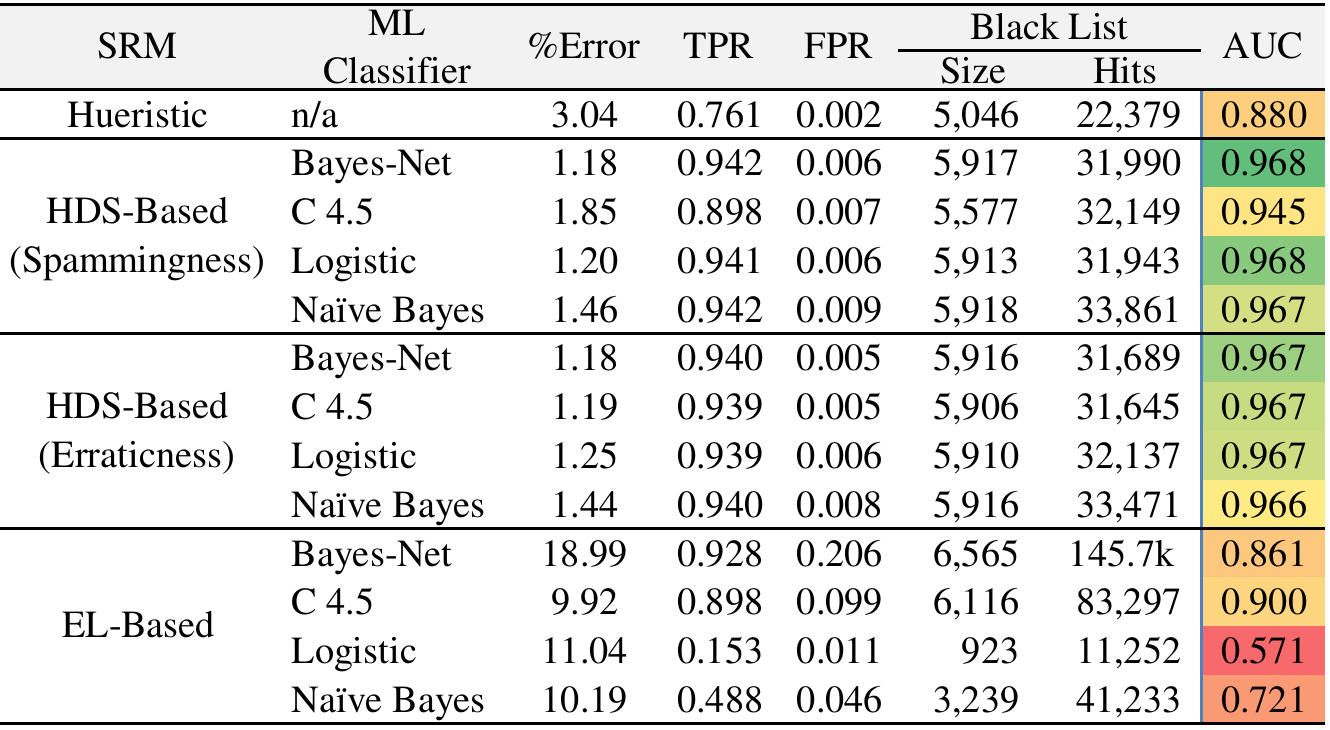}
\vspace{-0.3cm}
\end{table}

Judging the best results for each of the SRMs, we can see from Table \ref{tab:res-srm} that the HDS-SRM (Erraticness) had the best \(IP\) classification performance.  Not far behind at second place is the HDS-SRM (Spammingness). Third place, with noticeable AUC difference, is the EL-based SRM. The worst performance was achieved by the Heuristic-based SRM.
Interestingly, the best results for each SRM were obtained by different learning algorithms. Moreover, it seems that EL-based SRM is much more sensitive to the learning algorithm choice, compared with the HDS-Based SRMs. The Logistic-Regression, which worked very well for both HDS-SRMs, produced the worst results when applied to EL-Based. In this cases, very few \(IP\)s were blacklisted by the EL-based SRM, consequencly, obtaining a very poor AUC score.

The highest true-positive rate was acheived by the HDS-SRM (Spammingness), whereas the lowest false-positive and error rates were obtained by the HDS-SRM (Erraticness). The EL-based SRM blacklisted the most \(IP\)s. However, many of these  \(IP\)s were of benign senders, which is reflected in the very high false-positive rate obtained by this SRM.  
  
Finally, it is noticeable that in some configurations the very simple Heuristic and the EL-based SRMs achieved comparable performance.
We also noticed that both the Heuristic SRM and the EL-Based SRM roughly match the reported state-of-the-art performance.

\subsection{Batch IP Classification}
\label{sec:results-batch}

In the previous subsection we discussed continuous classification mode where the sender reputation is computed and updated each time a new email is received. This mode of operation may not be realistic due to relatively high resource consumption of machine learning based classifiers compared to black and white lists data structures. 
Moreover, classifying every incoming email would also mean placing the classifier in the critical path and turn it into a bottleneck during the process of handling incoming emails. 
Another drawback of the continuous classification mode is that most black-lists are optimized for fast information retrieval but do not tolerate frequent updates. 
Updating the black-list data structure may be a very expensive operation in terms of computational resources \cite{iplookup}.

It is therefore a good practice to minimize the number of updates and to make them as infrequent as possible.
In practice, SRM can be activated once in a while in order to save computational resources. 
The payoff for a periodic activation of SRM is a window of opportunity during which spammers who are not yet blacklisted can send large amounts of Spam without being blocked. 

In order to investigate the impact of black and white lists' update frequency on the accuracy of Spam filtering, we executed the reputation mechanisms in a batch mode with various update frequencies. 
The SRMs were executed each \(k\) minutes where \(k\) was set to 0.5, 1, 2, 5, 15, or 60 minutes.
HDS parameters were: \(w_0=60\) minutes, \(n=5\), and \(T_{Pred}=60\) minutes. 
In this experiment we used the BayesNet algorithm to train classifiers for both the HDS-Based SRMs and the EL-Based SRM. 

\begin{figure}[t]
\centering
\includegraphics[width=3.15in]{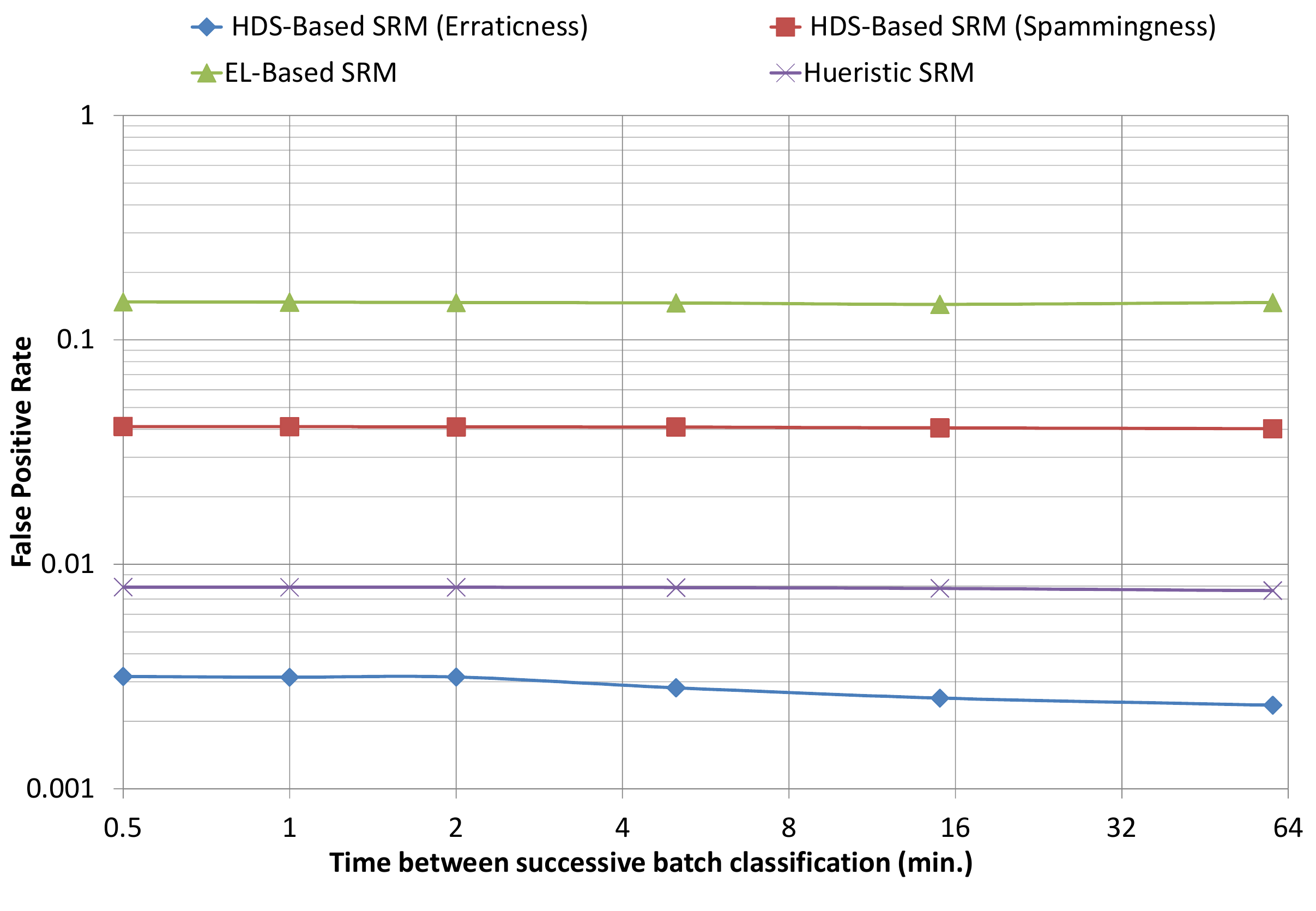}
\vspace{-0.3cm}
\caption{\label{fig:rslt-batch-fpr} FPR as a function of time between consequent address-lists updates.}
\vspace{-0.3cm}
\end{figure}

\begin{figure}[t]
\centering
\includegraphics[width=3.15in]{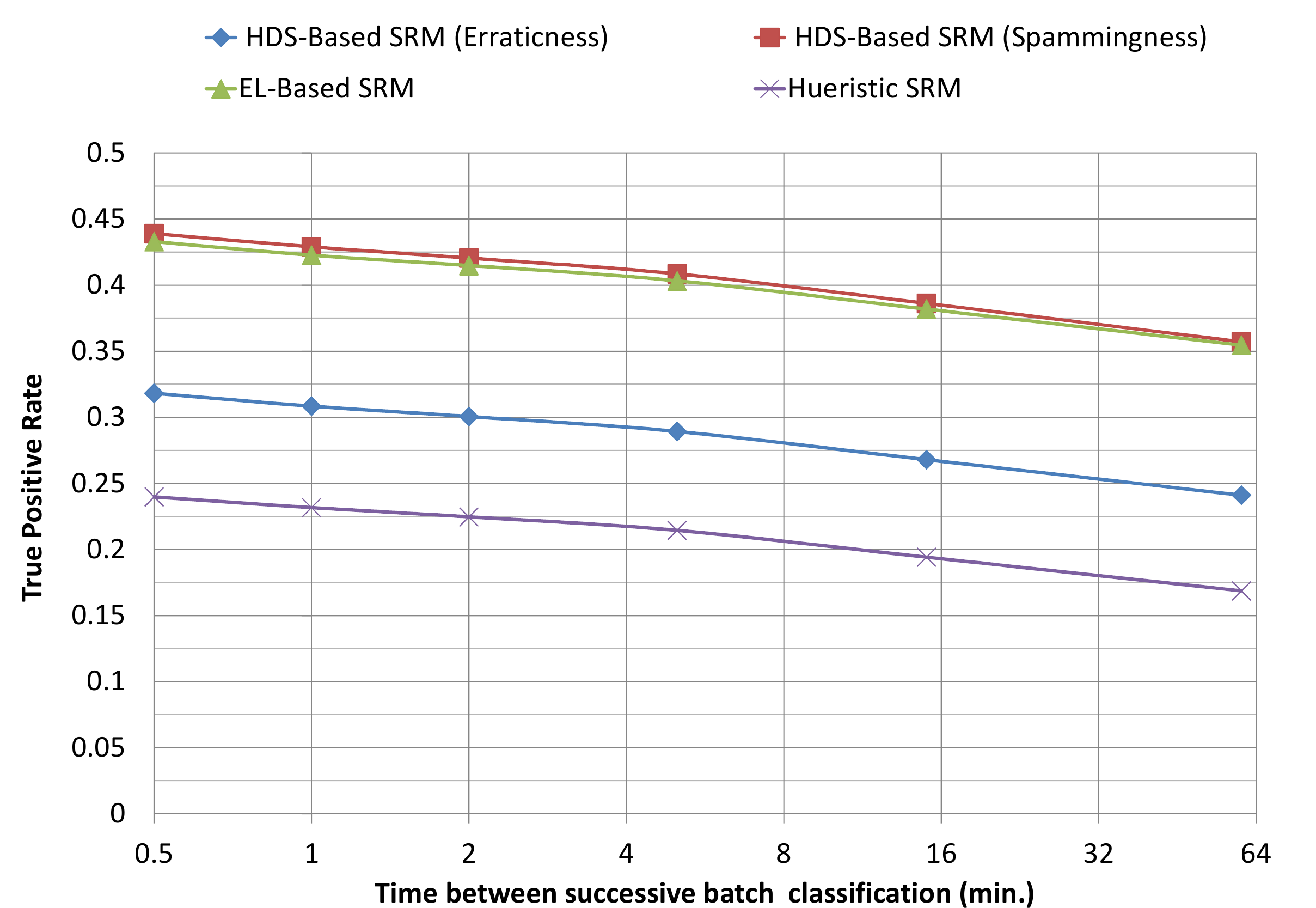}
\vspace{-0.3cm}
\caption{\label{fig:rslt-batch-tpr} TPR as a function of time between consequent address-lists updates.}
\vspace{-0.3cm}
\end{figure}

\begin{figure}[t]
\centering
\includegraphics[width=3.2in]{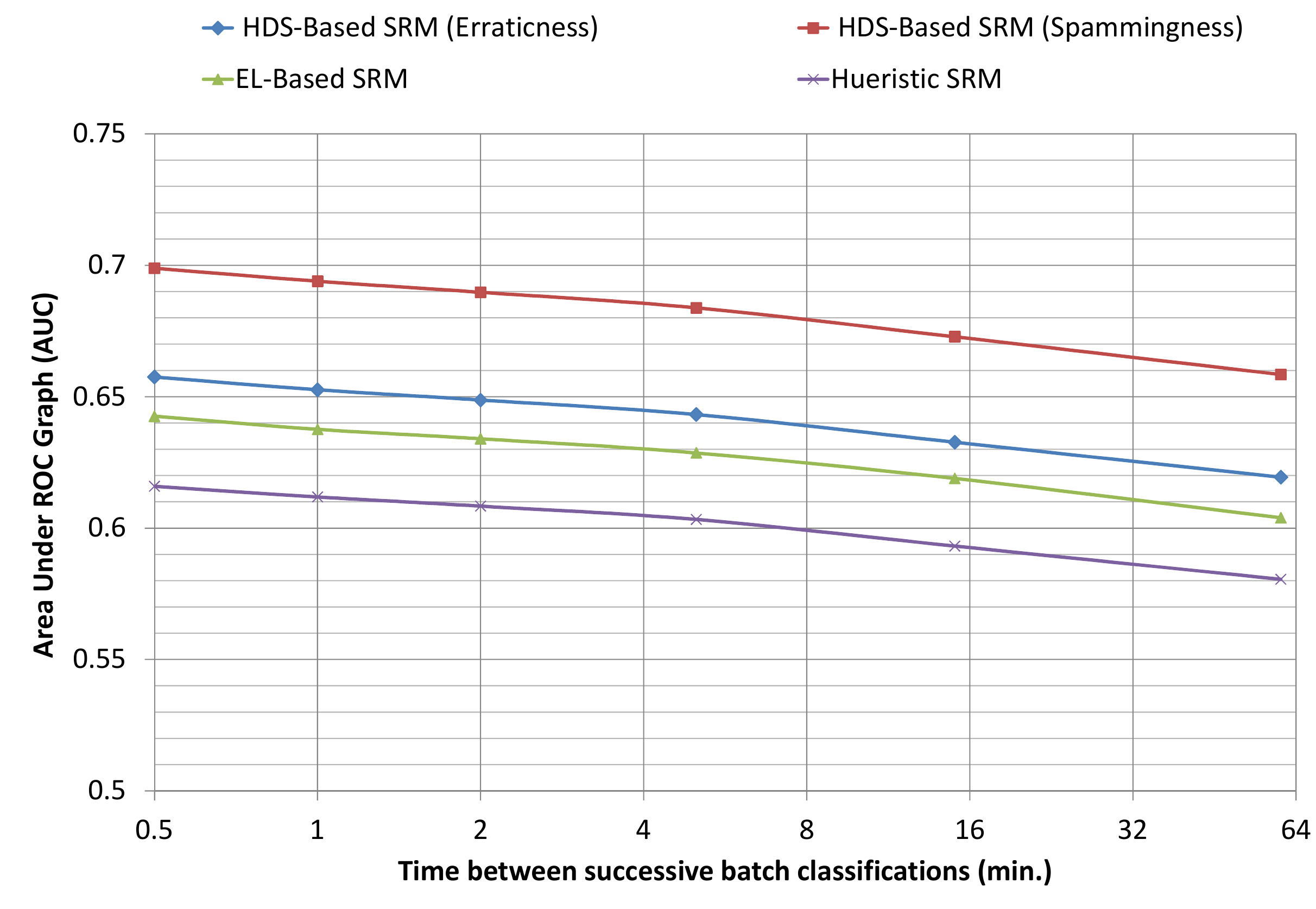}
\vspace{-0.3cm}
\caption{\label{fig:rslt-batch-auc} AUC as a function of time between consequent address-lists updates.}
\vspace{-0.3cm}
\end{figure}


The results presented in Figures \ref{fig:rslt-batch-fpr}, \ref{fig:rslt-batch-tpr}, and \ref{fig:rslt-batch-auc} depict the superiority of HDS-Based SRMs also in batch mode. 

Figure \ref{fig:rslt-batch-fpr} shows that the HDS-SRM (Erraticness) has the lowest false reject rate (FPR), while the EL-based SRM has the highest. Looking at the predictability results (TPR), we see that the Heuristic SRM had the worst results, whereas HDS-based SRM (Spammingness) had the highest results among the SRMs. Figure \ref{fig:rslt-batch-auc} shows that for all update frequencies, the HDS-SRM (Spammingness) has a considrebly high AUC when compared to the Heurist-based and El-based SRMs. Since the AUC metric is an objective performance metric that does not depend on a configurable threshold, it better reflects the superiority of the HDS over the other tested SRMs. 

In general, all four SRMs` performance detereorated, more or less at the same rate, as the black and white lists update frequency was reduced. This was well demonstrated by the drop in predictability, and AUC scores. Interestingly, the SRMs' false-positive rates were not sensitive to the update rate, and tend to stay constant.


\subsection{The Effect of History Length}
\label{sec:results-n}
The evidence presented in previous subsections suggests that aggregating sender MTA behavior over time is worthwhile and yields good classification models. 
The models created from the HDS when trying to predict the Spammingness of \(IP\) addresses are the least sensitive both to the choice of the machine learning algorithm and to the frequency of address-lists updates.
We therefore focus this subsection on the HDS-Based (Spammingness) SRM, investigating its performance as the function of the number of time windows used to construct the HDS.

We compared multiple models of the HDS-Based (Spammingness) SRM trained on different HDS train sets that were constructed from one to fourteen time windows. 
The models were induced by the Bayes-Net algorithm on HDS instances, generated for every incoming email using: \(w_0=15\) seconds, \(n=1,\ldots,14\) and \(T_{Pred}=60\) minutes. Both the black and white list were cleared once per 1,440 minutes (1 day).

\begin{figure}[t]
\centering
\includegraphics[width=3.15in]{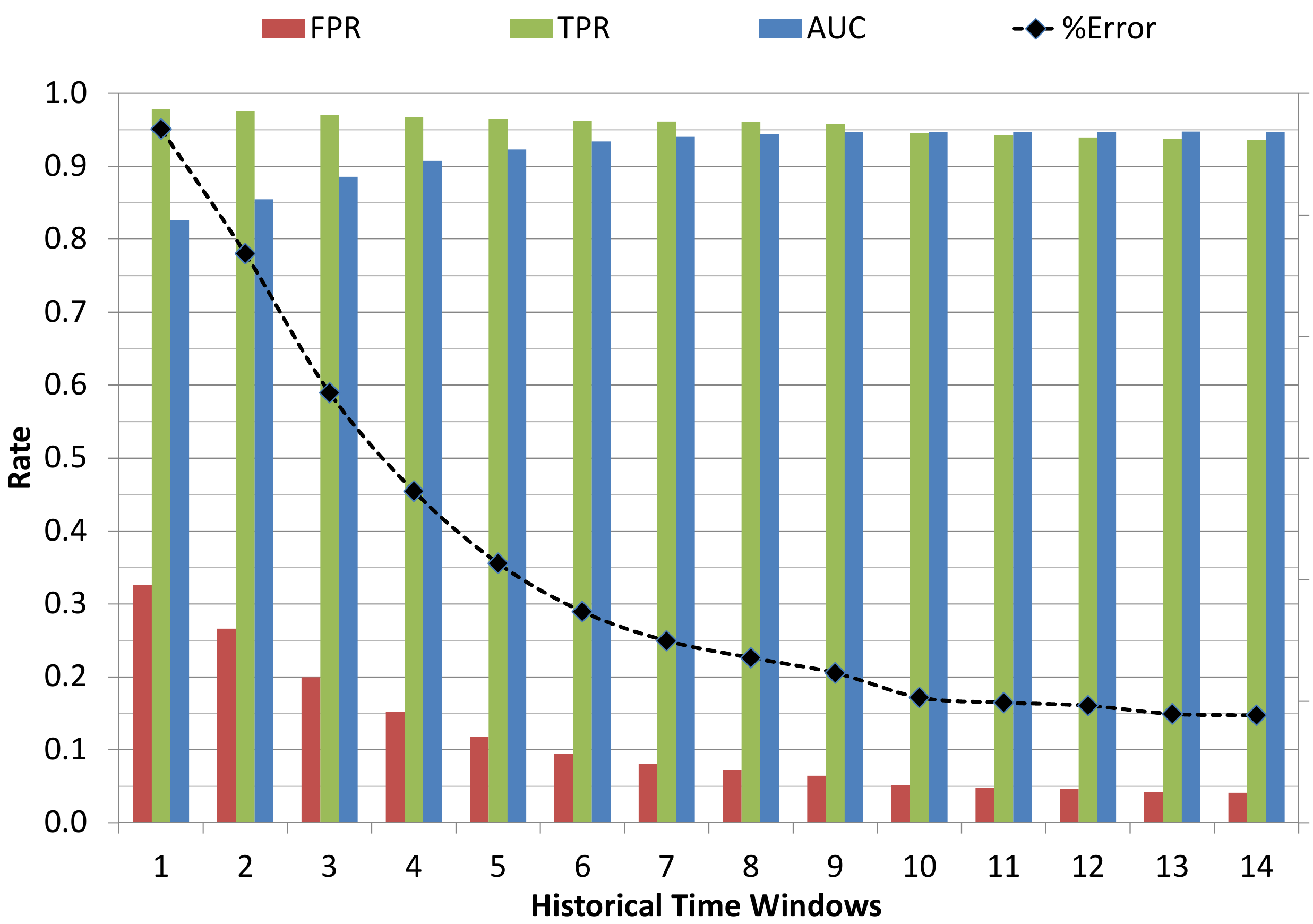}
\caption{\label{fig:res-history} The effect of history length on the SRMs` clssification performance}
\end{figure}

\begin{figure}[t]
\centering
\includegraphics[width=3.3in]{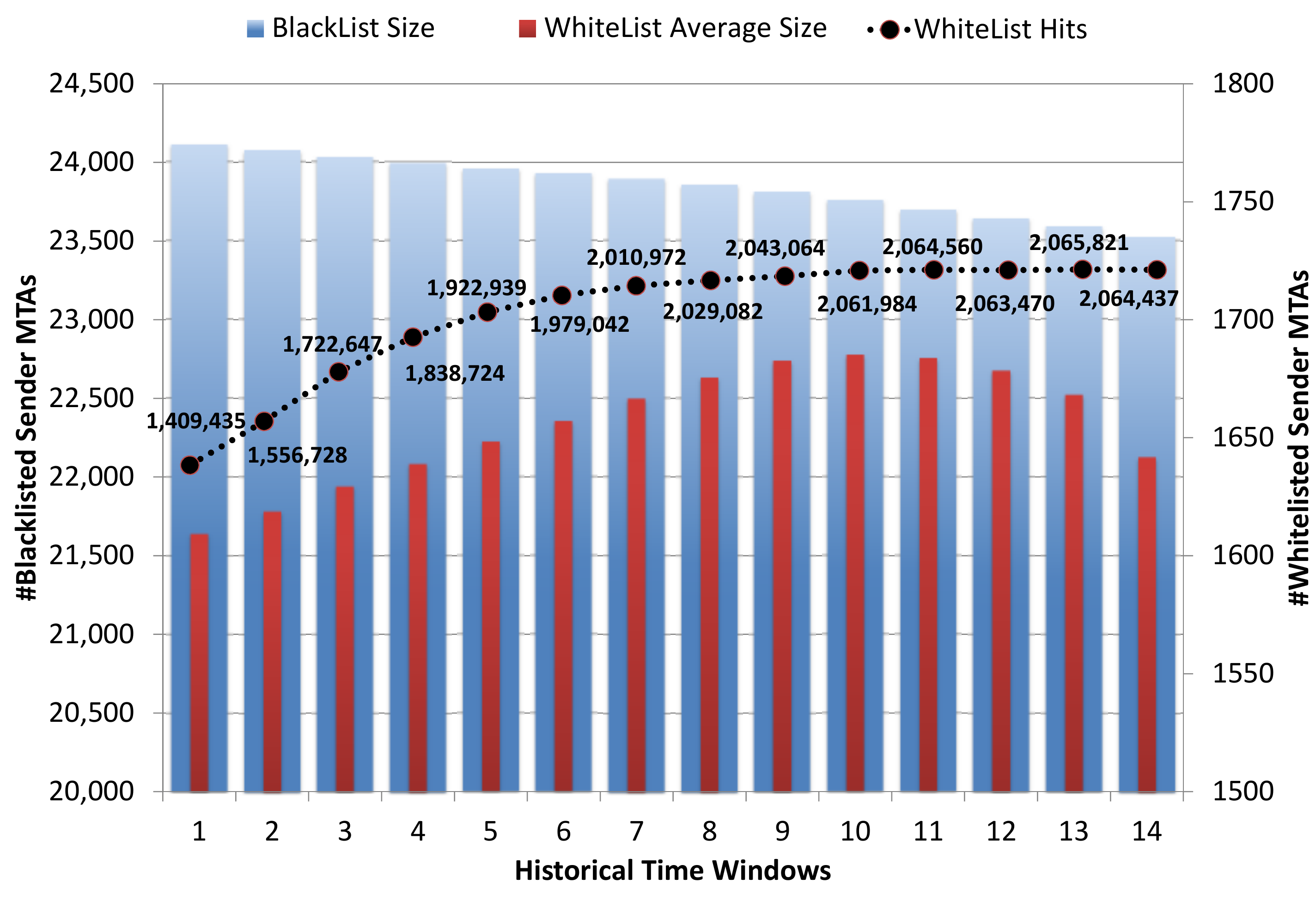}
\caption{\label{fig:res-history-BW-lists} The effect of history length on the black and white lists}
\end{figure}

\begin{table*}[bht]
\centering
\scriptsize
\caption{\label{tab:res-history} The effect of history length}
\begin{tabular}
{|l|@{\hspace{0.35mm}}c@{\hspace{0.35mm}}|@{\hspace{0.35mm}}c@{\hspace{0.35mm}}|@{\hspace{0.35mm}}c@{\hspace{0.35mm}}|@{\hspace{0.35mm}}c@{\hspace{0.35mm}}|@{\hspace{0.35mm}}c@{\hspace{0.35mm}}|@{\hspace{0.35mm}}c@{\hspace{0.35mm}}|@{\hspace{0.35mm}}c@{\hspace{0.35mm}}|@{\hspace{0.35mm}}c@{\hspace{0.35mm}}|@{\hspace{0.35mm}}c@{\hspace{0.35mm}}|@{\hspace{0.35mm}}c@{\hspace{0.35mm}}|@{\hspace{0.35mm}}c@{\hspace{0.35mm}}|@{\hspace{0.35mm}}c@{\hspace{0.35mm}}|@{\hspace{0.35mm}}c@{\hspace{0.35mm}}|@{\hspace{0.35mm}}c@{\hspace{0.35mm}}|}
\hline
Windows (\(n\)) & 1 & 2 & 3 & 4 & 5 & 6 & 7 & 8 & 9 & 10 & 11 & 12 & 13 & 14\\
\hline
Length (sec) 	& 15  		& 30 & 60 & 120 & 240 & 480 & 960 & 1,920 & 3,840 & 7,680 & 15,360 & 30,720 & 61,440 & 122,880\\
Error rate 		& 28.535 	& 23.407 & 17.683 & 13.639 & 10.673 & 8.683 & 7.484 & 6.783 & 6.163 & 5.152 & 4.941 & 4.816 & 4.474 & 4.420\\
FPR 					& 0.326 	& 0.266 & 0.199 & 0.152 & 0.118 & 0.094 & 0.080 & 0.072 & 0.065 & 0.051 & 0.048 & 0.046 & 0.042 & 0.041\\
TPR 					& 0.978 	& 0.976 & 0.970 & 0.967 & 0.964 & 0.962 & 0.961 & 0.961 & 0.958 & 0.945 & 0.942 & 0.940 & 0.937 & 0.935\\
BL size 			& 24,112 	& 24,074 & 24,031 & 23,993 & 23,958 & 23,928 & 23,892 & 23,855 & 23,810 & 23,758 & 23,698 & 23,640 & 23,589 & 23,523 \\
WL Hits 			& 1,409,435 & 1,556,728 & 1,722,647 & 1,838,724 & 1,922,939 & 1,979,042 & 2,010,972 & 2,029,082 & 2,043,064 & 2,061,984 & 2,064,560 & 2,063,470 & 2,065,821 & 2,064,437 \\
AUC 					& 0.826 & 0.855 & 0.885 & 0.907 & 0.923 & 0.934 & 0.940 & 0.944 & 0.947 & 0.947 & 0.947 & 0.947 & 0.948 & 0.947\\
\hline
\end{tabular}
\end{table*}

The experiment shows a mixed trend in the classification performance as a function of the number of historical time windows.
The results presented in Figures \ref{fig:res-history} and \ref{fig:res-history-BW-lists}, and Table \ref{tab:res-history} show that up to the ninth historical time-window the performance of the HDS-Based SRMs improves, whereas, when using more historical time-windows, the AUC score ceased to improve, indicating a change of trend.

Note that the experiment settings were not optimal w.r.t. initial historical time window (\(W_0=15s\)) resulting in lower TPR and higher FPR that the respective performance metrics reported in Section \ref{sec:results-el-hds}.
Yet, this experiment illustrates the effect of the number of time windows on performance of HDS-based SRM.
Due to the very small initial historical time window (i.e., only 15 seconds) we are able to notice the decrease in false positive and error rates as more time windows were used.
Surprisingly, the true positive rate was also gradually decreased as the number of historical time windows grew.
This can be explained by both the decrease in the blacklist size and the growing number of features that added additional dimensions to the machine-learning problems, and therefore, made it more and more complex to learn from (a.k.a. ``the course of dimensionality''). Notice that the FPR decreased faster then the TRP as the number of historical time windows increased.

The blacklist size declined at a constant rate, as more historical time windows were used. 
Note that the blacklist's size decline had probably affected (i.e., reduced) both the TPR and the FPR, since less \(IP\)s were classified as spammers.
In contrast to the black list, the white list average size increased until the tenth time window, and then the trend was reversed, were it begun an accelerated decline in size. 
Interestingly, the number of whitelist hits remained more or less stable from the 9th to the 14th time windows, even when the white list average size dropped. 
Since the AUC score during these time windows (9th to the 14th) was constant, we conclude that for the whitelisting task, the number of history time windows should be greater than nine. 
Overall, we see that as the more history is used beyond the 9th time window, the effectivess of the black and white lists increases. 


\section{Discussion and Future Work}
\label{sec:conclusions}
\subsection{IP Classification Performance}
Experiment results presented in this paper show that aggregating behavior of MTAs over time is an effective way to elicit valuable information from email logs.
The proposed method was found to be more effective than email-log-based and heuristic-based SRMs, tested under the \emph{exact same conditions}. 
In fact, the ISP whose dataset we evaluated uses a blacklisting method which is similar to the Heuristic SRM, while the Email-log-based SRM is similar to state-of-the-art methods \cite{behavior2, rep2, rep7}. 

The same machine learning algorithms applied on HDS produce much more effective models than if applied on the non-aggregated data extracted from the raw email logs.
The best results were obtained using HDS-Based (Erraticness) SRM with nine time windows: AUC 0.907, TPR 90.7\%, and FPR 0.4\%.
To the best of our knowledge these results are better than previously reported SRMs evaluated on data sets of similar scale.
In fact, the accuracy of HDS based SRMs approaches the accuracy of content-based filtering.
Another interesting fact is that HDS-based SRM blacklisted roughly the same number of \(IP\)s as the EL-based SRM, while incurring, by far, fewer classifications errors.

Some related works (e.g. \cite{rep2}) reported roughly the same performance as the EL-Based SRM reported here. 
Despite the differences in the particular features, we believe that aggregations over multiple time windows can boost the performance of sender reputation mechanisms based on statistical learning. 
ON our data set the upgrading of EL-based SRM to HDS-based SRM resulted in an elevated performance. We suggest that our results are general enough to motivate "upgrading" EL to HDS in other data set too.


In the second experiment we studied the impact of periodical execution of SRM on its effectiveness.
The results show a clear trade-off between batch size and the effectiveness of Spam filtering.
A less frequent execution of SRM results in less predictability power, as expected.
The main reason for inefficient blacklisting when sender reputation is computed once in a long time period is the tendency of spamming bots to send a number of Spam emails during a very short time period and go silent afterwards \cite{behavior1, survey2}. 
The observed deficiency of periodical activation of the evaluated SRMs could also be explained by a reduced marginal benefit when compared to the email service provider`s own SRM that is activated once in a while.

In Section \ref{sec:results-n} we studied the influence of history length on the performance of the reputation classifier. 
The results show that in general, the longer history is used the better classification model is produced. 

Increasing the number of time windows (and hence the number of features) above a certain point have resulted in an increasing efficiency of both the black and white lists, and thus a decreased workload of the entire filtering system. At the same time, instead of growing further, the AUC remained constant more or less. This phenomenon occurs probably due to the ``course of dimensionality'' effect, in which the growing dimensionality of the dataset plays a negative role, making the learned concept more and more complex.

In order to capture a longer mail sending history using fewer historical time windows, the size of the smallest historical time window \(w_0\)can be increased.
Unfortunately, in this case the most recent behavior of the MTAs would be diluted and damage the ability of the HDS-based SRM to respond to sudden behavioral changes. 
The ``course of dimensionality'' phenomenon can also be tackled by selecting the most informative features.
Currently, we leave the optimal configuration of HDS construction as an open issue.

Computing the reputation of a sender \(IP\) using HDS-based SRM is a computationally intensive task due to both the construction of HDS records and the classification using machine learning models. The HDS construction could be optimized by using past HDS records to compute the aggregated features of a new one. This should be further studied in a future work.
On the machine-learning end, for lowering the overall computational requirements of the HDS-based SRM we suggest using a non-complex classification models, e.g., Na\"{i}ve Bayes or Bayes-Net. 

\subsection{Reducing the Filtering Workload With HDS}
To insure that the end-users receive only very few spam-mails, ISPs usually employ a filtering mechanism based on black and white lists. These lists need to be updated frequently, so to insure minimal filtering errors. Currently, there are three methods for updateing these lists: real-time black listing (RBL), SRMs, and  content-based filters (CFBs). The SRMs and CBFs are much slower, and computation-power demanding, compared to the black and white lists. However, the CBF and SRMs only filter emails that were filtered-in by the black and white lists. Thus, as the black and white lists hit-ratio increase, fewer emails are need to be processed by the CBF and SRMs, and hence the filtering process becomes more computationally efficient. 
While a very high black and white list hit ratios might incur a very computentionally efficient filtering, it can result with high filtering errors, and so there is a constant trade-off between filtering efficiency and accuracy. 
In order to increase both the filtering computationally efficiency and accuracy, we propose using HDS-base SRMs, as method for updating both the black and white lists. 

On our experiments the black and the white lists had on average 457,120 and 1,916,636 hits respectively. Each hit corresponded to a spam or benign MTA's email that was not put through a content-based filter. Therefore, the filtering workload decrease (FGain) is:
\[F\-Gain=\frac{black\ and\ white\ lists\ hits}{emails\ in\ TestSet}=\] 
\[=\frac{2,373,756}{2,864,208}=0.828=82.7\%\]
This means that more than 4 out of 5 emails had hit one of the lists, and therefore, skipped the content-based filter, thanks to the HDS-RM black and white listing. This is a very significant contribution for ISPs, as their entire filtering workload, and conseqencly the energy consumed during the email filtering process can significanly be reduced.


\section{Acknowledgments}
This work was partially supported by Deutshce Telekom AG.
The authors would like to thank Yevgenia Gorodtzki, Igor Dvorkin, and Polina Zilberman for coding first versions of HDS and to Danny Hendler 
and Yuval Elovici for useful remarks.

\bibliographystyle{abbrv}
\bibliography{hds}

\begin{thebibliography}{10}

\bibitem{survey1}
Messaging anti-abuse working group (maawg) consumer survey part ii-detailed
  findings \& charts: A look at consumers` awareness of email security and
  practices., 2009.
\newblock
  \url{http://www.maawg.org/sites/maawg/files/news/2009_MAAWG-Consumer_Survey-Part2.pdf}.

\bibitem{rep1}
D.~Alperovitch, P.~Judge, and S.~Krasser.
\newblock Taxonomy of email reputation systems.
\newblock In {\em ICDCS Workshops'07}, 2007.

\bibitem{cbf1}
I.~Androutsopoulos, J.~Koutsias, K.~Chandrinos, and C.~D. Spyropoulos.
\newblock An experimental comparison of naive bayesian and keyword-based
  anti-spam filtering with personal e-mail messages.
\newblock In {\em ACM SIGIR}, pages 160--167. ACM Press, 2000.

\bibitem{rep4-sn}
J.~Balthrop, S.~Forrest, M.~E.~J. Newman, and M.~M. Williamson.
\newblock Technological networks and the spread of computer viruses.
\newblock {\em Science}, 304(5670):527--529, 2004.

\bibitem{rep7}
R.~Beverly and K.~Sollins.
\newblock Exploiting the transport-level characteristics of am.
\newblock In {\em 5th Conference on Email and Anti-Spam (CEAS)}, 2008.

\bibitem{cbf3}
J.~Blosser and D.~Josephsen.
\newblock Scalable centralized bayesian spam mitigation with bogofilter.
\newblock In {\em USENIX LISA}, pages 1--20, 2004.

\bibitem{rep5-sn}
P.~Boykin and V.~Roychowdhury.
\newblock Leveraging social networks to fight spam.
\newblock {\em IEEE Computer}, 38(4):61--68, 2005.

\bibitem{ml-Logistic-Regression}
S.~L. Cessie and J.~C.~V. Houwelingen.
\newblock Ridge estimators in logistic regression.
\newblock {\em Applied Statistics}, 41(1):191--201, 1992.

\bibitem{cbf2}
J.~Clark, I.~Koprinska, and J.~Poon.
\newblock A neural network based approach to automated email classification.
\newblock In {\em IEEE/WIC ýInternational Conference on Web Intelligence}, page
  702, 2003.

\bibitem{expurgate}
Eleven.
\newblock expurgate.
\newblock http://www.eleven.de/overview-antispam.html, June 2010.

\bibitem{rep6}
J.~Golbeck and J.~Hendler.
\newblock Reputation network analysis for email filtering.
\newblock In {\em First Conference on Email and Anti-Spam}, Mountain View,
  ýCalifornia, USA, 2004.

\bibitem{DBLP:journals/sigkdd/HallFHPRW09}
M.~Hall, E.~Frank, G.~Holmes, B.~Pfahringer, P.~Reutemann, and I.~H. Witten.
\newblock The weka data mining software: an update.
\newblock {\em SIGKDD Explorations}, 11(1):10--18, 2009.

\bibitem{rep2}
S.~Hao, N.~A. Syed, N.~Feamster, A.~G. Gray, and S.~Krasser.
\newblock Detecting spammers with snare: Spatiotemporal network-level automated
  reputation engine.
\newblock In {\em 18th USENIX Security Symposium}, 2009.

\bibitem{survey2}
K.~J. Higgins.
\newblock Botnets battle over turf.
\newblock http://www.darkreading.com/document.asp?doc\_id =122116, Apr. 2007.

\bibitem{ml-naiive-bayes}
G.~H. John and P.~Langley.
\newblock Estimating continuous distributions in bayesian classifiers.
\newblock In {\em Conference on Uncertainty in Artificial Intelligence}, pages
  338--345, San Mateo, 1995.

\bibitem{cbf4}
I.~Koprinska, J.~Poon, J.~Clark, and J.~Chan.
\newblock Learning to classify email.
\newblock {\em Inf. Sci.}, 177(10):2167--2187, 2007.

\bibitem{ipv6-spam}
P.~Kosik, P.~Ostrihon, and R.~Rajabiun.
\newblock Ipv6 and spam.
\newblock In {\em MIT Spam Conference}, 2009.

\bibitem{rep-oracle-based}
W.~Liu.
\newblock Identifying and addressing rogue servers in countering internet email
  misuse.
\newblock In {\em IEEE SADFE}, pages 13 --24, 2010.

\bibitem{TrustedSource}
McAfee.
\newblock Trustedsource.
\newblock http://www.trustedsource.org/.

\bibitem{survey3}
M.~Namestnikova.
\newblock Securelist spam report: December 2011.
\newblock
  \url{http://www.securelist.com/en/analysis/204792212/Spam_report_December_2011},
  Dec. 2011.

\bibitem{ml-c45}
J.~R. Quinlan.
\newblock {\em C4.5: programs for machine learning}.
\newblock Morgan Kaufmann Publishers Inc., San Francisco, CA, USA, 1993.

\bibitem{behavior1}
A.~Ramachandran and N.~Feamster.
\newblock Understanding the network-level behavior of spammers.
\newblock In {\em ACM SIGCOMM}, Pisa, Italy, 2006.

\bibitem{behavior2}
A.~Ramachandran, N.~Feamster, and S.~Vempala.
\newblock Filtering spam with behavioral blacklisting.
\newblock In {\em ACM CCS}, pages 342--351, 2007.

\bibitem{iplookup}
M.~A. Ruiz-Sanchez and E.~W. Biersack.
\newblock Survey and taxonomy of ip address lookup algorithms.
\newblock {\em IEEE Network}, 15(2):8--23, 2001.

\bibitem{rep-predictive-bl}
F.~Soldo, A.~Le, and A.~Markopoulou.
\newblock Predictive blacklisting as an implicit recommendation system.
\newblock In {\em IEEE INFOCOM}, pages 1 --9, 2010.

\bibitem{rbl3}
SORBS.
\newblock http://www.au.sorbs.net/.

\bibitem{rbl2}
SpamCop.
\newblock http://www.spamcop.net/bl.shtml.

\bibitem{rbl1}
Spamhaus.
\newblock http://www.spamhaus.org.

\bibitem{rep3}
Y.~Tang, S.~Krasser, P.~Judge, and Y.-Q. Zhang.
\newblock Fast and effective spam sender detection with granular svm on highly
  imbalanced mail server havior data.
\newblock In {\em 2nd International Conference on Collaborative Computing:
  Networking, Applications and Worksharing (CollaborateCom)}, Atlanta, Georgia,
  USA, 2006.

\bibitem{rep-spatio-temp}
A.~G. West, A.~J. Aviv, J.~Chang, and I.~Lee.
\newblock Preventing malicious behavior using spatio-temporal reputation.
\newblock In {\em ACM EUROSYS'10}, 2010.

\bibitem{weka}
I.~H. Witten and E.~Frank.
\newblock {\em Data Mining: Practical machine learning ýtools and techniques}.
\newblock Morgan Kaufmann, San Francisco, 2nd edition edition, 2005.

\bibitem{rep-collaborative-2}
M.~Xie and H.~Wang.
\newblock A collaboration-based autonomous reputation system for email
  services.
\newblock In {\em IEEE INFOCOM}, pages 1 --9, 2010.

\bibitem{cbf5}
S.~Youn and D.~McLeod.
\newblock Improved spam filtering by extraction of information from text
  embedded image email.
\newblock In {\em ACM SAC}, pages 1754--1755, New York, NY, USA, 2009.

\end{thebibliography}


\end{document}